\documentclass[]{spie}  %>>> use for US letter paper
\usepackage[]{graphicx}
\title{Design and characterization of TES bolometers and SQUID readout electronics for a balloon-borne application} 

\author{Johannes Hubmayr\supit{a}, Fran\c cois Aubin\supit{b}, Eric
Bissonnette\supit{b}, Matt Dobbs\supit{b}, Shaul Hanany\supit{a},
Adrian T. Lee\supit{c}, Kevin MacDermid\supit{b}, Xiaofan Meng\supit{c}, Ilan Sagiv\supit{a},
Graeme Smecher\supit{b} \skiplinehalf
\supit{a}University of Minnesota School of Physics and Astronomy, Minneapolis, MN 55455; \\
\supit{b}McGill University, Montr\'eal, Quebec, H3A 2T8, Canada; \\
\supit{c}University of California, Berkeley, Berkeley, CA 94720
}
\authorinfo{Contact Author: J.Hubmayr \\              Email: hubmayr@physics.umn.edu \\           Telephone: 612.626.9149}
\newcommand{\comment}[1]{}
\newcommand{\micron}[1]{$\mu$m}
\newcommand{\Gbar}[1]{$\bar{G}$}
\begin{document}
  \maketitle 

%%%%%%%%%%%%%%%%%%%%%%%%%%%%%%%%%%%%%%%%%%%%%%%%%%%%%%%%%%%%%
%%%%%%%%%%%%%%%%%%%%%%%%%%%%%%%%%%%%%%%%%%%%%%%%%%%%%%%%%%%%% 
\begin{abstract}

We present measurements of the electrical and thermal properties of new
arrays of bolometeric detectors that were fabricated as part of a
program to develop bolometers optimized for the low photon background of
the EBEX balloon-borne experiment.  An array consists of 140 spider-web
transition edge sensor bolometers microfabricated on a
4$\textquotedblright$ diameter silicon wafer.  The designed average
thermal conductance ($\bar{G}$) of bolometers on a proto-type array is
32~pW/K, and measurements are in good agreement with this value.  The
measurements are taken with newly developed, digital frequency domain
multiplexer SQUID readout electronics.

\end{abstract}


%%%%%%%%%%%%%%%%%%%%%%%%%%%%%%%%%%%%%%%%%%%%%%%%%%%%%%%%%%%%%
%%%%%%%%%%%%%%%%%%%%%%%%%%%%%%%%%%%%%%%%%%%%%%%%%%%%%%%%%%%%%
\section{INTRODUCTION}
\label{sec:intro}  % \label{} allows reference to this section

%----------------------------------------------------------------------

\comment{

Outline:

* Current science goals in astrophysics and cosmology require new levels
  of mm-wave telescope sensitivity
* Sensitivity achieved by large format arrays of detectors
* SQUID based multiplexing used to readout large arrays
* Systems currently in use are optimized for ground-based observation
         APEX-SZ - absorber coupled + analog mux
	 SPT - polarization sensitive + analog mux
	 ACT - PUD + time domain NIST
	 
* We optimize for balloon loading
* Low G, 10 pW/K
* Use DfMUX which consumes power at level suitable for balloon
}

%------------------------------------------------------------------------

Improvements in millimeter-wavelength instrument sensitivity are now
achievable through the development of large format bolometer arrays
cooled to sub-Kelvin temperatures.  Single bolometers have been
demonstrated to yield sensitivity close to the limit set by photon
noise.  Thus, increased sensitivity can only be achieved with large
numbers of detectors.  Superconducting transition edge sensor (TES)
bolometers are a suitable choice to make arrays because they can be
fabricated using standard thin film deposition and optical lithography
techniques~\cite{gildemeister:array}.  The TES bolometer also has the
advantage of strong negative electro-thermal feedback which increases
the linearity, dynamic range and speed of the device~\cite{lee:tes370}.
Readout electronics based on multiplexing superconducting quantum
interference devices (SQUIDs) are now capable of reading out large
arrays of TES bolometers.  Multiplexing is a key technology which
decreases the heat load on the sub-Kelvin detector stage as well as the
cost and complexity of cold wiring.  Time domain multiplexing has been
developed by NIST~\cite{deKorte:tdm} and an analog frequency domain
multiplexer (fMUX) system has been developed by a collaboration between
University of California, Berkeley (UCB) and Lawrence Berkeley National
Laboratory (LNBL)~\cite{lanting:fmux}.

Several ground-based experiments are now fielding these technologies
including APEX-SZ~\cite{dobbs:apex}, the South Pole Telescope
(SPT)~\cite{ruhl:spt} and the Atacama Cosmology Telescope
(ACT)~\cite{fowler:act} to probe the Cosmic Microwave Background (CMB)
radiation.  Observation at frequencies above 250 GHz are difficult from
the ground and can provide critical information for CMB experiments such
as characterizing the dust foreground.  We are building a balloon-borne
experiment called EBEX~\cite{oxley:ebex} that is designed to measure the
polarization of the CMB and will implement 1440 TES bolometers with
SQUID based multiplexed readouts. EBEX will have three frequency bands
centered around 150, 250 and 410~GHz, each with $\sim$~30\% bandwidth.

The lower atmospheric loading and colder telescope temperature at
balloon altitudes allow substantial sensitivity gains by lowering the
thermal conductance of the bolometer. The designed average thermal
conductance ($\bar{G}$) of the EBEX 150~GHz bolometers is 10~pW/K.  The
designed optical time constant is 3~ms.

In this paper we report measurements of a prototype array of bolometers
that was fabricated as a step in achieving the low thermal conductance
necessary for EBEX. This array was designed to have $\bar{G}=32$~pW/K
and a predicted time constant of 3~ms.

The measurements were performed with digital frequency domain
multiplexer (DfMUX) electronics developed at McGill
University~\cite{dobbs:dfmux}.  The DfMUX system is an upgrade from the
analog fMUX system and was specifically designed for the low power
consumption requirements of a balloon experiment.

%%%%%%%%%%%%%%%%%%%%%%%%%%%%%%%%%%%%%%%%%%%%%%%%%%%%%%%%%%%%%%%%%%%%
%%%%%%%%%%%%%%%%%%%%%%%%%%%%%%%%%%%%%%%%%%%%%%%%%%%%%%%%%%%%%%%%%%%%
\section{Bolometer Array}
\label{sec:bolometer}

%------------------------------------------------------------------------
\comment{
Points:

* wafer
  * # per wafer
  * center to center spacing
  * overall size 
  * quarter lambda backshort (bonded wafer): 250 nm backshort
  * leads to wirebond
* Bolo design 
  * Nitride height: 1um
  * leg length: 1mm, .5 mm
  * width (2 legs:15 um, 6 legs: 6 um)
  * Au web height: 12 nm
  * Au web width: 2um
  * Au bling height: 80 nm
  * Au bling ring width: 32 um
  * 2.1 mm diameter absorber
  * grid spacing (117 um)
  * leg length and width (1 mm, 3 um wide (except leads))
  * metalization
  * G
  * tau opt
* TES specs
  * Al/Ti prox. effect bilayer
  * Size
  * Rnormal
  * Tc
  * Bling
* Fab techniques
  * 
}

%------------------------------------------------------------------------

\begin{figure}[t]
   \begin{center}
   \begin{tabular}{c}
   \includegraphics{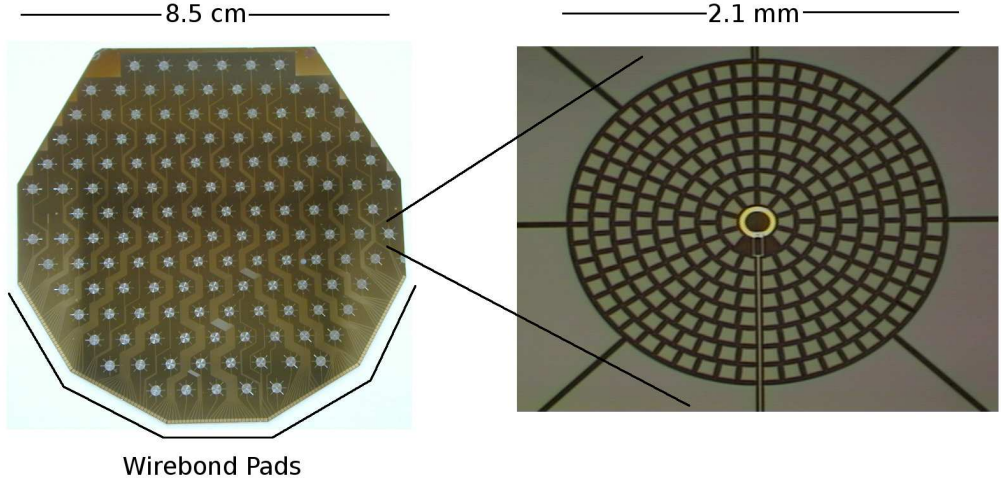}
   \end{tabular}
   \end{center}
   \caption[example] 
%>>>> use \label inside caption to get Fig. number with \ref{}
   { \label{fig:wafer} {\bf Left:} The 140 element transition edge
   sensor (TES) bolometer array.  {\bf Right:} A close up picture of a
   spider-web TES bolometer.  The gold ring and TES can be seen in the
   middle of the picture with the superconducting leads exiting the
   bottom of the picture.}
   \end{figure}

The left panel of Fig.~\ref{fig:wafer} shows a photograph of the
prototype bolometer array, which was fabricated in the Berkeley Microlab
clean-room facility using standard thin film deposition and optical
lithography.  The array contains 140 spider-web TES bolometers spaced
6.6~mm apart center-to-center.  Superconducting aluminum leads connect
each TES to wirebonding pads at the bottom five sides of the wafer.  One
of the bolometers is shown in the right panel of Fig.~\ref{fig:wafer}.
The bolometer consists of three main structures: a spider-web absorber,
a TES and a gold ring.  The 2.1~mm diameter spider-web absorber is
composed of 1~$\mu$m thick, 6~$\mu$m wide low-stress silicon nitride and
has a 117~$\mu$m grid spacing.  The spider web geometry is chosen to
reduce heat capacity as well as the cross-section to cosmic rays.  The
web is metallized with a 2~$\mu$m wide, 12~nm thick layer of Au which
has a DC sheet resistance $\sim$~200~$\Omega$ per square.  The
spider-web is thermally isolated from the heatsink by silicon nitride
legs that have a ratio of cross-sectional area to length $A/l = 132$~nm.
The transition edge sensor is composed of an Al/Ti proximity effect
sandwich tuned to have a $\sim$~1~$\Omega$ normal resistance and
transition temperature $T_c \sim$~500 mK.  The sensor is thermally
attached to a gold ring.  The heat capacity of the gold ring limits the
sensor bandwidth ensuring stability~\cite{irwin:tesstability}.  The
backside of the 500~$\mu$m thick wafer is coated with a 250~nm layer of
gold creating a $1/4\lambda$ backshort at 150~GHz, near the peak of the
CMB spectrum.

%--------------
%-------------

We predict the dynamic thermal conductance of the bolometers using
$G=4\sigma A T^3 \xi$~\cite{holmes:nitride}, where
$\sigma$=15.7mW/cm$^2$K$^4$ and $T$ is the temperature of the TES.  We
use the complete diffuse surface scattering limit known as the Casimir
limit $\xi \approx \sqrt{A}/l$ which is valid for $\sqrt{A}/l \ll$ 1.
We predict $G$ = 61~pW/K at $T_{c}$.  The relationship between the
average thermal conductance ($\bar{G}$) and the dynamic thermal
conductance ($G$) is~\cite{lee:tes100}
\begin{equation}
	\label{eq:G}
G/\bar{G} = \frac{(n+1)(1-T_s/T_c)}{1-(T_s/T_c)^{n+1}}
	\end{equation}
assuming the thermal conductivity follows $\kappa\sim T^{n}$.  Heat
conduction from the bolometer to heatsink is dominated by silicon
nitride, which has an index $n=3$.  Using Eqn.~\ref{eq:G} we predict
$\bar{G}$ = 32~pW/K.

%%%%%%%%%%%%%%%%%%%%%%%%%%%%%%%%%%%%%%%%%%%%%%%%%%%%%%%%%%%%%%%%%%%%%%%%%%%%%%%%%%%%%%%
%%%%%%%%%%%%%%%%%%%%%%%%%%%%%%%%%%%%%%%%%%%%%%%%%%%%%%%%%%%%%%%%%%%%%%%%%%%%%%%%%%%%%%%

\section{Readout} 
\label{sec:expsetup} 
\subsection{Principle}
\label{sec:subsec:principle}

\begin{figure}[t]
   \begin{center}
   \begin{tabular}{c}
   \includegraphics[height=2in,width=3.75in]{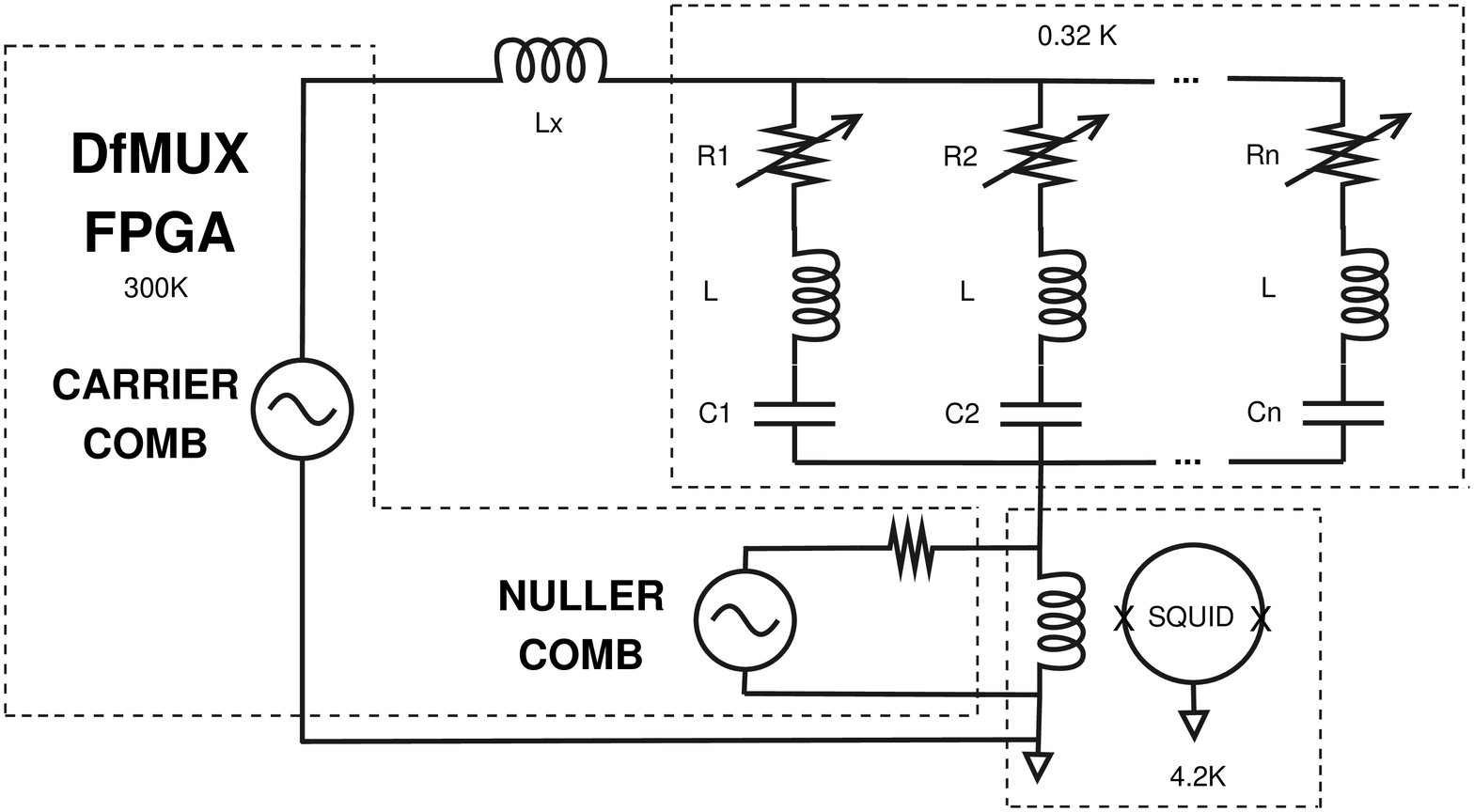}
   \end{tabular}
   \end{center}
   \caption[example] 
%>>>> use \label inside caption to get Fig. number with \ref{}
   { \label{fig:schematic} Electrical schematic of the readout system.
     The multiplexed module is held at 320~mK.  The bolometers in the
     module are voltage biased with a comb of sine wave carriers.
     Currents from the bolometers are read out with a SQUID ammeter at
     4.2~K.  The carrier amplitudes are removed with a nuller comb.
     The carrier and nuller combs are produced with an FPGA on the
     DfMUX readout boards described in Sec.~\ref{sec:subsec:dfmux}.
     $L_x$ is the stray inductance in series with the module discussed
     in Sec.~\ref{sec:subsec:netanal}.}
   \end{figure}

We use SQUID based frequency domain multiplexing to readout the
bolometer arrays.  An electrical schematic of the readout system is
shown in Fig.~\ref{fig:schematic}.  The $\sim$~1~$\Omega$ transition
edge sensors of the bolometer array are placed in series with band
defining LC filters.  The LCR circuits are wired in parallel to create
a multiplexed module.  A comb of sine wave carriers between
300~kHz~-~1~MHz voltage biases each sensor in the module at its LC
resonant frequency.  Sky intensity changes the sensor resistance and
amplitude modulates the carrier transferring signals to the side-bands
of each carrier.  Thus each sensor response is well defined in
frequency space.  Currents from all sensors within the module are
carried on a single pair of wires to a SQUID ammeter.  Warm
electronics are used to lock in on the carrier frequency of each
sensor with a bandwidth that contains the sky signal.

The large carrier amplitudes of the module present a flux burden on
the SQUID.  Since there is no sky signal at the carrier frequencies,
they can be removed.  A nuller comb consisting of sine wave currents
180 degrees out of phase with each carrier frequency is summed at the
SQUID input.  This nuller comb removes the unmodulated carrier
amplitudes reducing the dynamic range requirement of the SQUID and
increasing linearity.

%% tabular environment useful for creating an array of images  
%-------------
%-------------  

\subsection{DfMUX}
\label{sec:subsec:dfmux}

\begin{figure}[t]
    \begin{center}
   \begin{tabular}{c}
   \includegraphics[height=2.in]{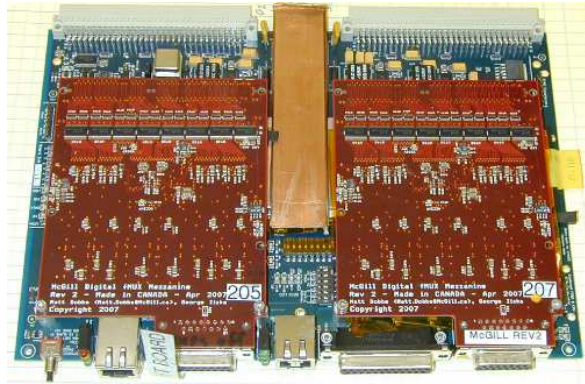}
   \end{tabular}
     \end{center}
   \caption[example] 
%>>>> use \label inside caption to get Fig. number with \ref{}
   { \label{fig:dfmux} The digital frequency domain multiplexer
   (DfMUX) readout board developed at McGill University is shown.  Two
   mezzanine boards which house the majority of the analog components
   attach to the FPGA motherboard.  One board can readout 128
   bolometers and dissipates 16W.}
   \end{figure}

We use new digital frequency domain multiplexer (DfMUX) electronics for
our implementation of frequency domain multiplexing.  The DfMUX boards
are a drop-in replacement for the analog fMUX boards designed to consume
substantially less power and improve low frequency noise
performance~\cite{dobbs:dfmux}.  The DfMUX board shown in
Fig.~\ref{fig:dfmux} produces the sine wave carrier and nuller combs
digitally with a Xilinx Virtex4 LX160 FPGA.  The SQUID output is
directly digitized with an ADC operating at 25~MHz.  Sky signal from
each bolometer is then digitally demodulated with a set of parallel
algorithms that use a half-wave mixer and a series of cascading filters.
As such the mixer is sensitive to odd harmonics of the fundamental
frequency with a response that falls of as $1/(2n+1)$, where $n$ in an
integer.

%--------------
%-------------

Placing sine wave generation and signal demodulation in the digital
domain achieves substantial power savings over the previous
implementation of the frequency multiplexing system, which performed
these functions in the analog domain.  The power consumption of one
DfMUX board is 16~W and is capable of reading out four modules with
multiplexing factors of 32.  The DfMUX system is therefore capable of
dissipating as little as 125 mW per detector.  However, the
multiplexer bandwidth is currently limited by the cold components of
the system.  Using a multiplexing factor of 12 to readout 1440
detectors, as is planned for EBEX, the power consumption of the
readout system is 565 W, which includes a 15\% loss of efficiency in
power delivery.  This level of power consumption satisfies the EBEX
power budget constraints.

The DfMUX boards contain a set of algorithms to tune and monitor the
SQUIDs and detectors.  This functionality is essential for commanding
the readout system remotely during a balloon flight.  The measurements
described in Sec. \ref{sec:results} are taken with automated scripts run
by the DfMUX boards.

\section{Experimental Setup}
\label{sec:subsec:setup}

The prototype bolometer array is heatsunk to the baseplate of a
${^3}$He adsorption refrigerator~\cite{chase:fridge} operated at
320~mK.  The array is enclosed in a dark cavity at the same
temperature so that radiative loading is negligible.  A Fairchild
LED56 inside the dark cavity provides optical signals to the
bolometers.  Each bolometer is wired in series with a ceramic
capacitor~\cite{panasonic} and a 16~$\mu H$ inductor fabricated by
TRW.  Using a multiplexing factor of three, the bolometers are read
out with a 100 series array SQUID amplifier~\cite{huber:squid}
operated in shunt feedback coupled to a room temperature amplifier
located on a custom SQUID Controller electronics
board~\cite{spieler:sqcontroller}.  The SQUID is heatsunk to 4.2~K.
The output of the SQUID is sent on a twisted pair to the demodulator
of the DfMUX board.

%%%%%%%%%%%%%%%%%%%%%%%%%%%%%%%%%%%%%%%%%%%%%%%%%%%%%%%%%%%%%%%%%%%%
%%%%%%%%%%%%%%%%%%%%%%%%%%%%%%%%%%%%%%%%%%%%%%%%%%%%%%%%%%%%%%%%%%%%
\section{Results}
\label{sec:results} 
\subsection{Network Analysis}
\label{sec:subsec:netanal} 
\begin{figure}[t]
    \begin{center}
   \begin{tabular}{c}
    \includegraphics[height=5in,angle=270]{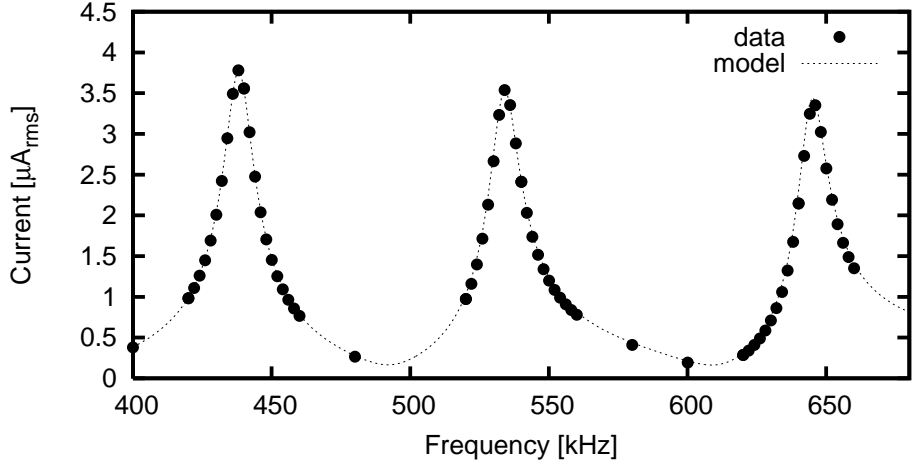}
   \end{tabular}
     \end{center}
   \caption[example] 
%>>>> use \label inside caption to get Fig. number with \ref{}
   { \label{fig:networkanalysis} The network analysis of the multiplexed
    module.  A fit to the analytic model shown by the dashed curve is
    used to determine the bias frequencies, resistance of the bolometers
    and the stray inductance in series with the module.}
   \end{figure}

In order to determine the bias frequencies,
normal resistance of the bolometers and the stray inductance in series
with the module, we perform the network analysis shown in
Fig.~\ref{fig:networkanalysis}.
%--------------
%-------------
For this measurement, the cold stage is held above the TES transition at
800 mK, and a carrier voltage sweeps across the module in frequency.
The three peaks at 438, 534 and 645 kHz are the LC resonant frequencies
in series with sensors `11-01', `10-02' and `9-03' respectively.  We fit
the data to an analytic circuit model with $2n+1$ free parameters where
$n$ is the number of bolometer channels.  The inductors in series with
each bolometer are fixed at 16~$\mu H$, and the voltage bias is fixed at
3.9~$\mu V_{rms}$.  For each channel the normal resistance of the
bolometer and the capacitance in series with the bolometer are to be
determined from the fit.  The stray inductance in series with the module
($L_x$, as shown in Fig.~\ref{fig:schematic}) is also determined by the
fit.  The impedance of the stray inductance ($\omega L_x$) creates a
frequency dependant voltage divider which is the source of the
decreasing peak height with increasing frequency.  Since the voltage
bias of the bolometer ($V_b$) largely determines the responsivity of the
device~\cite{lee:tes100}, it is important to determine $L_x$.  From the
fit, we determine that $R$ = 1.03, 1.04 and 1.00 $\Omega$ for bolometers
`11-01', `10-02' and `9-03' respectively, and $L_x$ = 149 nH.
\comment{The sum of squares of residuals is 0.2161 $\mu A_{rms}^2$ for
64 degrees of freedom.}  For all TES, $L_x$ affects the voltage bias by
$<$ 10\%.  However, for subsequent measurements, we include $L_x$ = 149
nH in our analysis.

\subsection{Thermal Conductance}
\label{sec:subsec:iv}

\begin{figure}[b]
    \begin{center}
   \begin{tabular}{c}
   \includegraphics[height=5in,angle=270]{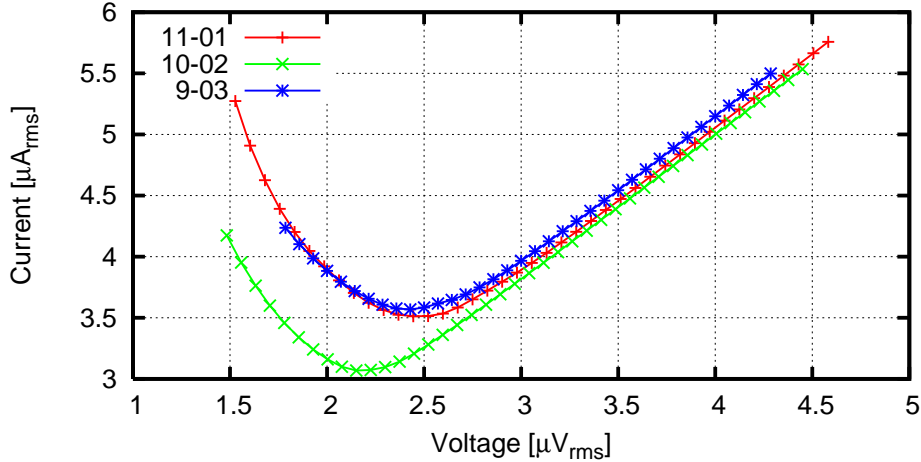}
   \end{tabular}
     \end{center}
   \caption[example] 
%>>>> use \label inside caption to get Fig. number with \ref{}
   { \label{fig:iv} Current versus voltage curves for the bolometers
     in the module.  The electrical power at the turnaround divided by
     $T_c-T_s$ gives $\bar{G}$ = 32, 27 and 33~pW/K for the three
     devices.}
   \end{figure}

To determine the average thermal conductance of the bolometers
($\bar{G}$), we perform current versus voltage (IV) measurements of
the three bolometers in the module.  Each sensor is biased at its
resonant frequency, the voltage is stepped down and the current
through the SQUID is recorded.  At voltage biases $> 3 \mu V_{rms}$
shown in Fig.~\ref{fig:iv}, the TES is normal and the IV curve is
linear.  The turnaround at $\sim 2.5 \mu V_{rms}$ is evidence that the
TES enters the superconducting transition.  In the transition, the
total power is constant due to strong electro-thermal feedback, and the
current is proportional to the inverse of the voltage bias.  The
steady state power through the device is
 \begin{equation}
\label{eqn:steadystate}
 P_{rad} + P_{elect} = \bar{G}(T_{tes}-T_s),
\end{equation}  
where $P_{rad}$ is the radiative power, $P_{elect}=V_b^2/R_{tes}$ is
the electrical power, $\bar{G}$ is the average thermal conductance,
$T_{tes}$ is the temperature of the TES and $T_s$ is the temperature
of the heatsink.  Since the bolometers are operated within a dark
enclosure, the data in Fig. \ref{fig:iv} together with $T_c - T_s$
yield measurements of $\bar{G}$.  We determine $T_c \sim$~550~mK by
biasing the bolometers with $\sim$~10~$nV_{rms}$ and monitoring the
current response while slowly lowering the heatsink temperature.  We
measure $\sim$ 32, 27 and 33 pW/K for bolometers `11-01', `10-02' and
`9-03' respectively, which are in good agreement with the
theoretically calculated $\bar{G}$ = 32~pW/K.

%--------------
%-------------

\subsection{Optical frequency response}
\label{sec:subsec:tauopt} 

\begin{figure}[t]
    \begin{center}
   \begin{tabular}{c}
   \includegraphics[height=5in,angle=270]{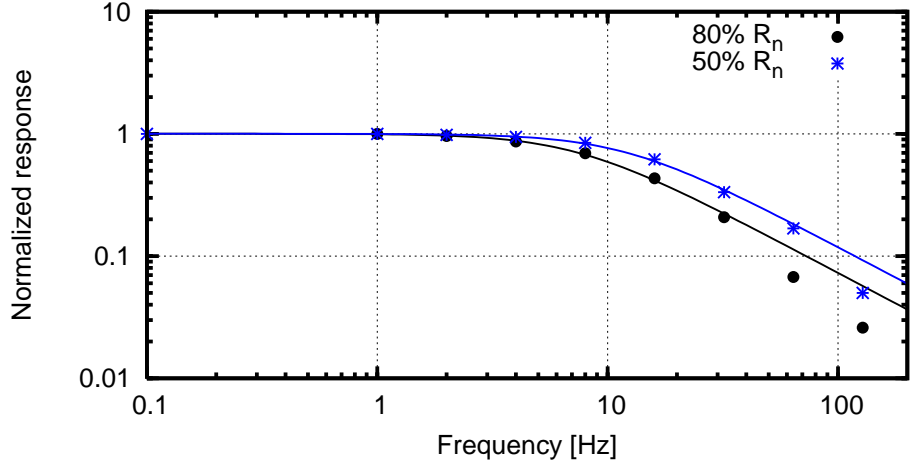}
   \end{tabular}
     \end{center}
   \caption[example] 
%>>>> use \label inside caption to get Fig. number with \ref{}
   { \label{fig:tau} The optical frequency response of bolometer
   `9-03' biased to 0.8~$\Omega$ (black dots) and 0.5~$\Omega$ (blue
   asterisks).  The single pole fits yield 22~ms and 13~ms time
   constants.}
   \end{figure}

We determine the bolometer response to optical signals by biasing the
LED with a small sinusoidal current and measuring the bolometer
amplitude response as a function of LED bias frequency.  Figure
\ref{fig:tau} shows the frequency response of bolometer `9-03' biased at
0.8~$\Omega$ (black circles) and 0.5~$\Omega$ (blue asterisks).  A
single-pole fit gives optical time constants of 22~ms and 13~ms
respectively.  \comment{The sum of squares of residuals for the fits are
0.0048 and 0.0030 for seven degrees of freedom for each.}With feedhorns
coupled to the bolometers we expect the response time to decrease by a
factor of $\sim$ 2-3, which yields a response time close to our design
goals when biased at 0.5~$\Omega$.

The decreased time constant lower into the transition is evidence that
the thermalization of the TES, not the spider-web, dominates the
response time of the bolometer.  For bolometer arrays currently in
fabrication we have reduced the heat capacity of the gold ring by a
factor of four, which should decrease the response time of the TES by
the same factor.  The optical response time of the bolometer should
then be limited by the thermalization time of the web.

\comment{points at high frequency not a good fit.  Fit to a two-pole
model which you know is a more physically motivated model.}  

%--------------
%-------------
  
\comment{
Noise overview:

white noise consistent with thermal noise sources
1/f knee 200 mHz
Biased at 80% Rn
SQUID noise delemma
No added noise from adjacent channels
write down noise budget
comment on improving noise?
DfMUX noise negligible}

\subsection{Bolometer Noise}
\label{sec:subsec:noise}

\begin{figure} 
    \begin{center}
   \begin{tabular}{c}
   \includegraphics[height=5in,angle=270]{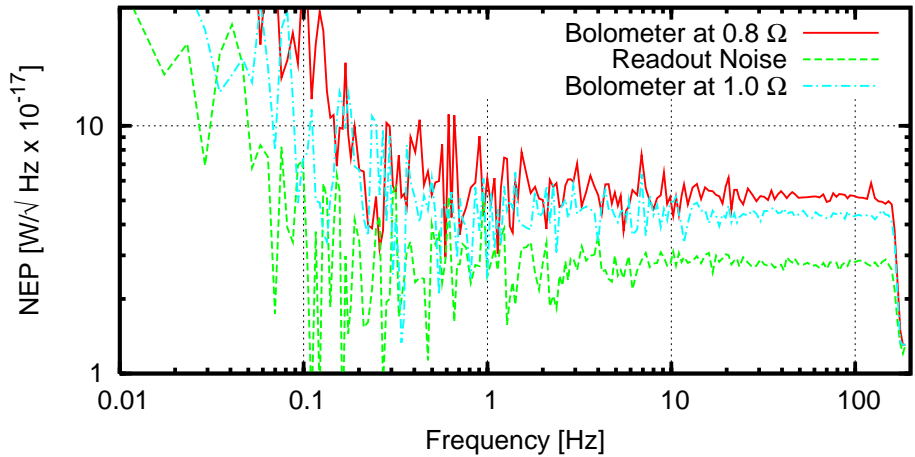}
   \end{tabular}
     \end{center}
   \caption[example] 
%>>>> use \label inside caption to get Fig. number with \ref{}
   { \label{fig:noise} Demodulated noise of bolometer `9-03' in NEP
    units biased into the transition (red, solid) and above the
    transition (blue, dot-dash).  The green, dashed curve shows the
    readout noise level.}
   \end{figure}

The demodulated noise spectrum of bolometer `9-03' is shown in
Fig. \ref{fig:noise}. The solid, red curve shows the noise level of the
bolometer when it is biased with 2.275~$\mu V_{rms}$ and has a
resistance of 0.8~$\Omega$.  The spectrum is white down to 200~mHz with
an amplitude of $5.0 \times 10^{-17}\,\, W/\sqrt{Hz}$.  The blue,
dot-dashed spectrum shows the noise of the bolometer when biased above
the transition at 1.0~$\Omega$ and is therefore insensitive to phonon
noise.  For this bias position the expected noise sources are bolometer
Johnson, SQUID and readout electronics noise.  The readout noise level
is shown in the green, dashed curve.  Readout noise consists of SQUID
noise and readout electronics noise.  When biased to 0.8~$\Omega$ the
expected noise level is 4.2~$\times 10^{-17} W/\sqrt{Hz}$, which is
calculated from the quadrature sum of readout, bolometer Johnson and
phonon noise using the measured thermal conductance value.  The 20\%
discrepancy between the calculated and measured noise levels is
currently under investigation.

%--------------
%-------------

The expected noise sources and levels for bolometer `9-03' are listed in
Tab.~\ref{tab:noise} in NEP units.  The demodulator transfer function is
different for signal and the different noise components of the system.
To account for these differences a factor of $\pi/2$ has been included
for the SQUID and readout electronics noise levels, and a factor of
$\sqrt{2}$ has been included for the bolometer Johnson noise.

\begin{table}[b]
\caption{
\label{tab:noise} 
Noise expectation for dark bolometer `9-03.'}
\begin{center}       
\begin{tabular}{l c c c} %% this creates two columns
%% use of \rule[]{}{} below opens up each row

\hline
\rule[-1ex]{0pt}{3.5ex}  Noise source & Equation & NEP \\ 
\rule[-1ex]{0pt}{3.5ex}  &  & ($10^{-17} W/\sqrt{Hz}$) \\
\hline\hline
\rule[-1ex]{0pt}{3.5ex}  Phonon~\cite{mather:noise} & $\sqrt{\gamma 4k_bT_c^2G}$ & 2.3 \\
%\hline
\rule[-1ex]{0pt}{3.5ex} Bolometer Johnson & $\sqrt{4 k_b T_c/R} \cdot
V_b$ & 1.9 \\
%\hline
\rule[-1ex]{0pt}{3.5ex}  SQUID & 2.5 $pA_{rms}/\sqrt{Hz} \cdot
 V_{b}$ & 0.9\\
%\hline
\rule[-1ex]{0pt}{3.5ex} Readout electronics & 4.7 $pA_{rms}/\sqrt{Hz}
 \cdot V_{b}$ & 1.7\\ 
\hline 
\end{tabular}
\end{center}
\rule[-1ex]{55pt}{0ex} Here $\gamma$ = 0.498, $k_b$ is Boltzmann's constant, $T_c$ = 550
mK, $G$ = 63 pW/K, $R$ = 0.8 $\Omega$ \\ \rule[-1ex]{55pt}{0ex} and $V_b$ = 2.275 $\mu
V_{rms}$.
\end{table}

Post-demodulation the readout noise is increased by a factor 1.5 due to
the half-wave mixer's sensitivity to odd harmonics of the demodulator
frequency. A measurement of the pre-demodulated noise level shows an
increase in noise at frequencies between 1 and 10 MHz. Since the mixer
is sensitive to odd harmonics of the fundamental it samples this excess
noise, which then adds to the demodulated noise level. The quadrature
sum of the mixer's response and the measured noise levels at odd
harmonics of the demodulator frequency is 3.0~$\times 10^{-17}
W/\sqrt{Hz}$ which matches the measured value.  The excess demodulated
noise can be addressed without hardware changes by filtering away
harmonics with a low pass filter in firmware.  The source of increased
out of band noise above 1~MHz is currently unknown, but we have ruled
out the bolometers as the source because the same out of band noise is
observed in SQUIDs that are not connected to detectors.

Bolometer Johnson, SQUID and readout electronics NEP scale linearly
with the voltage bias, $V_b$.  In any astrophysical application, the
radiative background loads the bolometer, and the voltage bias
required to keep the TES in transition is smaller than the voltage
bias needed for these dark measurements.  If half the power required
to keep these bolometers in the transition comes from radiative
loading, the NEP from terms proportional to voltage bias summed in
quadrature is 1.0 $\times 10^{-17} W/\sqrt{Hz}$ which is below the
photon noise level at 150~GHz of 2.5~$\times 10^{-17} W/\sqrt{Hz}$.

\comment{For half power calculation.  total noise level is 3.6.  G noise
2.2.  Noise sources other than photons 2.6.  therefore not photon noise
dominated.  3.45 pW optical power assumed at 150 GHz.}

%%%%%%%%%%%%%%%%%%%%%%%%%%%%%%%%%%%%%%%%%%%%%%%%%%%%%%%%%%%%%%%%%%%%
%%%%%%%%%%%%%%%%%%%%%%%%%%%%%%%%%%%%%%%%%%%%%%%%%%%%%%%%%%%%%%%%%%%%
\section{Conclusion}
\label{sec:conclusion} 

We have fabricated and measured a TES bolometer array as part of a
program to produce low thermal conductance TES arrays for the EBEX
balloon-borne experiment.  Average thermal conductance measurements of
three bolometers on the proto-type array are in good agreement with the
32~pW/K designed value.  Noise measurements are 20\% larger than
expected.  All measurements are taken with DfMUX readout electronics,
which have been designed for low power consumption suitable for a
balloon application.

With its design sensitivity and a 14 day flight, EBEX will
either detect the signature of gravity waves from the epoch of
inflation shortly after the big bang or will set a $2\sigma$ upper
bound of $1.3 \times 10^{16}$ GeV on the energy scale at which
inflation took place.  The bolometer arrays and DfMUX readout boards
will be tested in the balloon environment during the EBEX North
American flight scheduled for Fall 2008.

%%%%%%%%%%%%%%%%%%%%%%%%%%%%%%%%%%%%%%%%%%%%%%%%%%%%%%%%%%%%%%%%%%%%
\subsection{Acknowledgments} 

EBEX is supported by NASA through grant numbers NNG05GO02H and
NNX08AG40G. J. Hubmayr acknowledges support from the NASA Graduate
Student Research Program (GSRP) and a Grant-In-Aid of Research (GIAR)
from the National Academy of Sciences, administered by Sigma Xi, The
Scientific Research Society. We thank K.~Irwin and G.~Hilton for
providing SQUID arrays.  M. Dobbs acknowledges the support of the
Natural Sciences and Engineering Research Council of Canada (NSERC)
and of the Canadian Institute for Advanced Research (CIfAR) through
its Cosmology and Gravity Program.

%%%%%%%%%%%%%%%%%%%%%%%%%%%%%%%%%%%%%%%%%%%%%%%%%%%%
\appendix    %>>>> this command starts appendixes
%%%%%%%%%%%%%%%%%%%%%%%%%%%%%%%%%%%%%%%%%%%%%%%%%%%%

%%%%%%%%%%%%%%%%%%%%%%%%%%%%%%%%%%%%%%%%%%%%%%%%%%%%%%%%%%%%%
%%%%% References %%%%%

\bibliography{mybib}   %>>>> bibliography data in mybib.bib

\begin{thebibliography}{10}

\bibitem{gildemeister:array}
Gildemeister, J.~M., Lee, A.~T., and Richards, P.~L., ``Monolithic arrays of
  absorber-coupled voltage-biased superconducting bolometer,'' {\em Appl. Phys.
  Lett}~{\bf 77}(24),  4040--4042 (2000).

\bibitem{lee:tes370}
Lee, S., Gildemeister, J.~M., Holmes, W., Lee, A.~T., and Richards, P.~L.,
  ``Voltage-biased superconducting transition-edge bolometer with strong
  electrothermal feedback operated at 370 m{K},'' {\em Appl. Opt.}~{\bf 37},
  3391--3397 (1998).

\bibitem{deKorte:tdm}
de~Korte, P. A.~J., Beyer, J., Deiker, S., Hilton, G.~C., Irwin, K.~D.,
  MacIntosh, M., Nam, S.~W., Reintsema, C.~D., and Vale, L.~R., ``Time-division
  superconducting quantum interference device multiplexer for transition-edge
  sensors,'' {\em Rev. Sci. Instrum.}~{\bf 74},  3087 (2003).

\bibitem{lanting:fmux}
Lanting, T.~M., Cho, H., Clarke, J., Dobbs, M., Lee, A.~T., Richards, P.~L.,
  Smith, A.~D., and Spieler, H.~G., ``A frequency-domain squid multiplexer for
  arrays of transition-edge superconducting sensors,'' {\em IEEE Trans. Appl.
  Sup.}~{\bf 13}(2),  626 (2003).

\bibitem{dobbs:apex}
Dobbs, M. and N.{~}Halverson{~}{\it et al.}, ``{APEX-SZ} first-light and
  instrument status,'' {\em New Astronomy Reviews}~{\bf 50},  960--968 (2006).

\bibitem{ruhl:spt}
J.{~}Ruhl{~}{\it et al.}, ``The {S}outh {P}ole {T}elescope,'' {\em Proc. SPIE
  Int. Soc. Opt. Eng.}~{\bf 5543} (2004).

\bibitem{fowler:act}
J.{~}Fowler{~}{\it et al.}, ``The atacama cosmology telescope project,'' {\em
  Proc. SPIE mm. \& Sub-mm. Det. Ast. II}~{\bf 5498},  1--10 (2004).

\bibitem{oxley:ebex}
P.{~}Oxley{~}{\it et al.}, ``The {EBEX} experiment,'' {\em Proc. SPIE Int. Soc.
  Opt. Eng.}~{\bf 5543},  320--331 (2004).

\bibitem{dobbs:dfmux}
Dobbs, M., Bissonnette, E., and Spieler, H., ``Digital frequency domain
  multiplexer for mm-wavelength telescopes,'' {\em IEEE Transactions on Nuclear
  Science}~{\bf TNS-00230-2007.R2} (2008).

\bibitem{irwin:tesstability}
Irwin, K.~D., Hilton, G.~C., Wollman, D.~A., and Martinis, J.~M.,
  ``Thermal-response time of superconducting transition-edge
  microcalorimeters,'' {\em J. Appl. Phys.}~{\bf 83}(8),  3978--3985 (1998).

\bibitem{holmes:nitride}
Holmes, W., Gildemeister, J.~M., and Richards, P.~L., ``Measurements of thermal
  transport in low stress silicon nitride films,'' {\em Appl. Phys. Lett}~{\bf
  72},  2250--2252 (1998).

\bibitem{lee:tes100}
Lee, A.~T., Richards, P.~L., Nam, S.~W., Cabrera, B., and Irwin, K.~D., ``A
  superconducting bolometer with strong electrothermal feedback,'' {\em Appl.
  Phys. Lett}~{\bf 69}(12),  1801--1803 (1996).

\bibitem{chase:fridge}
Chase, S.,  [{\em Two-stage sub-Kelvin {$^3He$}
  cooler}{\nolinebreak\hspace{0.1em}]}, Chase {R}esearch {C}ryogenics Ltd., 140
  Manchester Road, Sheffield S10 5DL, England.

\bibitem{panasonic}
Panasonic,  [{\em ECJ series}{\nolinebreak\hspace{0.1em}]}.

\bibitem{huber:squid}
Huber, M.~E., Neil, P.~A., Benson, R.~G., Burns, D.~A., Corey, A.~M., Flynn,
  C.~S., Y.Kitaygorodskaya, Massihzadeh, O., Martinis, J.~M., and Hilton,
  G.~C., ``{DC} {SQUID} series array amplifiers with 120 {MH}z bandwidth
  (corrected),'' {\em IEEE Trans. Appl. Sup.}~{\bf 11}(2),  4048--4053 (2001).

\bibitem{spieler:sqcontroller}
Spieler, H., ``Frequency domain multiplexing for large scale bolometer
  arrays,'' {\em Monterey Far-IR, Sub-mm and mm Detector Technology Workshop
  proceedings} ,  243--249 (2002).

\bibitem{mather:noise}
Mather, J., ``Bolometer noise: nonequilibrium theory,'' {\em Appl. Opt.}~{\bf
  21}(6),  1125--1129 (1982).

\end{thebibliography}
\bibliographystyle{spiebib}   %>>>> makes bibtex use spiebib.bst
\end{document}